\documentclass[aps,twocolumn,showpacs,superscriptaddress,amssymb]{revtex4}
\date{\today}

\usepackage{graphicx}

\begin{document}

\title{Enhanced Stability of Superheavy Nuclei due to High-Spin Isomerism}

\author{F.R. Xu}
\affiliation{School of Physics, Peking University, Beijing 100871, China}
\affiliation{Institute of Theoretical Physics, Chinese Academy of Sciences,
Beijing 100080, China}
\affiliation{Center for Theoretical Nuclear Physics, National Laboratory
for Heavy Ion Physics, Lanzhou 730000, China}
\author{E.G. Zhao}
\affiliation{Institute of Theoretical Physics, Chinese Academy of Sciences,
Beijing 100080, China}
\affiliation{Center for Theoretical Nuclear Physics, National Laboratory
for Heavy Ion Physics, Lanzhou 730000, China}
\author{R. Wyss}
\affiliation{AlbaNova University Centre, KTH (Royal Institute of Technology),
S-104 05 Stockholm, Sweden}
\author{P.M. Walker}
\affiliation{Department of Physics, University of Surrey, Guildford,
Surrey GU2 7XH, UK}

\begin{abstract}
Configuration-constrained calculations of potential-energy surfaces 
in even-even superheavy nuclei reveal systematically the existence 
at low excitation energies of multi-quasiparticle states with deformed
axially symmetric shapes and large angular momenta.
These results indicate the prevalence of long-lived, 
multi-quasiparticle isomers. In a quantal system, the ground state is usually 
more stable than the excited states. In contrast, in superheavy nuclei
the multi-qausiparticle excitations decrease the probability for
both fission and $\alpha$ decay, implying enhanced stability. Hence,
the systematic occurrence of multi-qausiparticle isomers 
may become crucial for future production and study of even heavier
nuclei. The energies of multi-quasiparticle states and their $\alpha$ decays
are calculated and compared to available data.
\end{abstract}


\maketitle 

It has long been a fundamental question as to what are the maximum charge and
mass that a nucleus may attain. Recently, significant progress has been made
experimentally in the synthesis of the heaviest elements \cite{Hofmann00,
Oganessian99}. According to classical physics, elements with $Z\geq 104$
should not exist due to the large Coulomb repulsive force. The occurrence
of superheavy elements with $Z\geq 104$ is entirely due to quantal shell
effects. Theoretically, predictions have been made that there should be an
island of stability of superheavy nuclei around $N=184$ and $Z=114$, 120
or 126, depending on the model employed \cite{Nilsson68,Cwiok96,Smol97,
Rutz97,Bender99}.

Information concerning the excited-state structure of superheavy nuclei 
is scarce, so that the testing of structure calculations has been 
limited. A promising area of progress is represented by the observation of
rotational $\gamma$-ray transitions in $^{252,254}_{102}$No
\cite{Herzberg01,Reiter99} and $^{254,256}_{100}$Fm \cite{Fire96},
showing the deformed character of these nuclei, which can be described
theoretically \cite{Sob01}. The $\gamma$-ray transitions from non-collective 
one-quasiparticle states form another important probe into the structure of 
superheavy nuclei \cite{Cwiok99}. By measuring the properties and decays
of quasiparticle states, essential information about the nucleonic orbits
close to the Fermi surfaces can be gained. Indeed, detailed knowledge of
single-particle states is required for the development of reliable nuclear
structure models.

In the present work we consider multi-quasiparticle (multi-qp) states
formed by breaking pairs of nucleons. Unpaired nucleons couple their
angular momenta to states with total spin projection $K$ onto the symmetry
axis in deformed, axially symmetric nuclei. The $K$ values can be large, and,
due to the high degree of their $K$ forbiddenness, $\gamma$-ray transitions
from high-$K$ into low-$K$ states are strongly hindered, leading to the
formation of long-lived high-$K$ isomers. In addition, there are influences
on fission and $\alpha$-particle emission, which can be important in
increasing the survival probability of superheavy nuclei. Indeed, high-$K$
isomers might become crucial in the exploration of even heavier nuclei.
In the present work, we focus on high-spin isomerism and the angular-momentum
effect on the stability of even-even superheavy nuclei.

Configuration-constrained potential-energy-surface (PES) calculations
\cite{Xu98} have been performed to determine the deformations and
excitation energies of multi-qp states. Single-particle levels are obtained
from the non-axial deformed Woods-Saxon potential with the set of
universal parameters \cite{Naz85}. In order to reduce the unphysical
fluctuation of the weakened pairing field (due to the blocking effect of
unpaired nucleons) an approximate particle-number projection has been used
by means of the Lipkin-Nogami method \cite{Sat94}
with pairing strengths determined by the average gap method \cite{Mol92}.
In the configuration-constrained PES calculation, it is required to
adiabatically block the unpaired nucleon orbits which specify a given
configuration. This has been achieved by calculating and identifying
the average Nilsson quantum numbers for every orbit involved in a
configuration \cite{Xu98}. The total energy of a state consists of a
macroscopic part that is obtained with the standard liquid-drop model
\cite{Mye66} and a microscopic part which is calculated by the Strutinsky
shell-correction approach, including blocking effects. The PES is calculated
in the space of quadrupole ($\beta_2$, $\gamma$) and hexadecapole
($\beta_4$) deformations. The configuration-constrained PES calculation
can properly treat the shape polarization due to unpaired nucleons.

\begin{figure}
\includegraphics[scale=0.45]{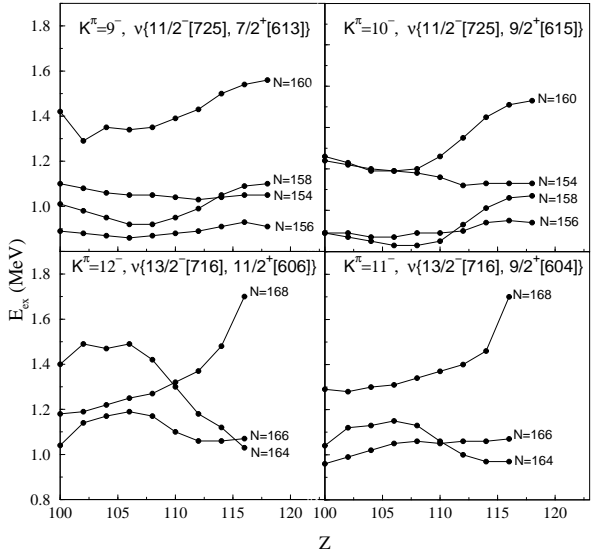}
\caption{\label{figure1}Calculated excitation energies for two-quasineutron
states.}
\end{figure}

\begin{figure}
\includegraphics[scale=0.45]{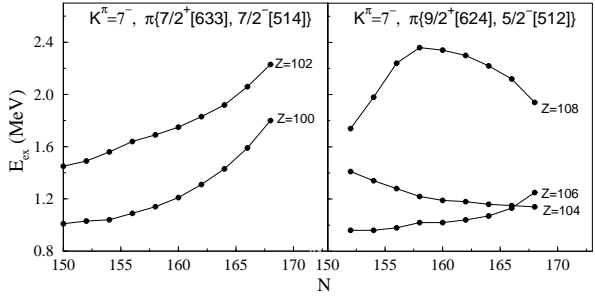}
\caption{\label{figure2}Calculated excitation energies for two-quasiproton
states.}
\end{figure}

Following predictions from theory, many experiments
\cite{Herzberg01,Reiter99,Fire96} have demonstrated the systematic
existence of deformed superheavy nuclei. In fact, almost all the superheavy
nuclei found experimentally are believed to be deformed, and
the stability of superheavy nuclei is enhanced due to deformed shell effects. 
Furthermore, deformed high-$K$ orbits close to the Fermi surfaces
can form various high-$K$ multi-qp states at low excitation energies.
In the present work, we are interested in possible isomeric states.
The combination of high $K$, low energy and axially deformed shape provides 
the necessary conditions for the formation of isomers. Additionally, a parity 
opposite to that of the ground state (g.s.) can further reduce the transition
rate from an excited state to the g.s., or to a member of the g.s.
band (g.s.b.). Figs.~\ref{figure1} and \ref{figure2} display our calculations
of low-energy high-$K$ two-qp states with negative parity for even-even
$Z\geq 100$  nuclei. These investigated states have deformed, axially
symmetric equilibrium shapes. 

\begin{figure}
\includegraphics[scale=0.45]{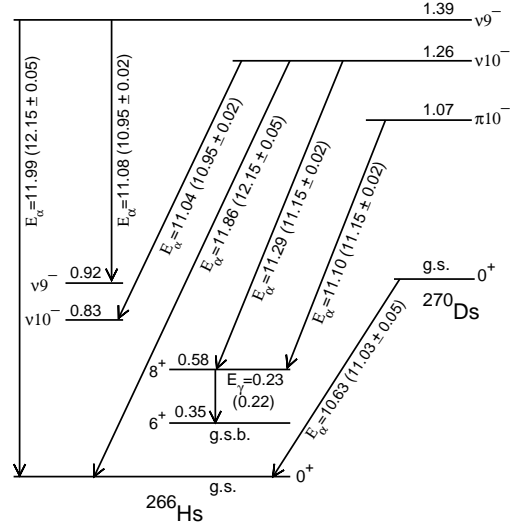}
\caption{\label{figure3}Schematic figure for the $\alpha$ decays corresponding
to the reported decay events \cite{Hofmann01} from $^{270}$Ds to $^{266}$Hs.
Different options for the initial isomer are indicated. The energies
(in MeV) are not drawn to scale. The configurations are: 
$\nu 9^- = \nu\frac{11}{2}^-[725]\otimes\frac{7}{2}^+[613]$; 
$\nu 10^- = \nu\frac{11}{2}^-[725]\otimes\frac{9}{2}^+[615]$;
$\pi 10^- = \pi\frac{11}{2}^+[615]\otimes\frac{9}{2}^-[505]$.
The values in parentheses are the experimental energies \cite{Hofmann01}.}
\end{figure}

In a recent experiment \cite{Hofmann01}, $\alpha$ decays from isomeric
states in $^{270}_{110}$Ds were observed with a half-life of
$6.0^{+8.2}_{-2.2}$ ms that is significantly longer than the
$100^{+140}_{-40}$ $\mu$s of the corresponding g.s. Possible configurations
for these isomers were proposed to be
$K^\pi=9^-$ ($\nu\frac{11}{2}^-[725]\otimes\frac{7}{2}^+[613]$) and/or
$10^-$ ($\nu\frac{11}{2}^-[725]\otimes\frac{9}{2}^+[615]$) \cite{Hofmann01}. 
Our calculations confirm these assignments and indeed show that
the $9^-$ and $10^-$ two-quasineutron states exist systematically at
low excitation energies (see Fig.~\ref{figure1}) with axially symmetric
deformations around $\beta_2\sim 0.2$.  In $^{270}$Ds, the calculated
energies of the $9^-$ and $10^-$ isomers are 1.39 and 1.26 MeV, respectively,
agreeing with the experimental estimate of 1.13 MeV \cite{Hofmann01}.
Furthermore, configuration-constrained PES calculations allow us to
determine the $Q_\alpha$ value of the $\alpha$ decay from a given
configuration of the parent nucleus to a given state of the daughter.
Then, the $\alpha$-particle kinetic energy is obtained from
$E_\alpha=\frac{A_p-4}{A_p}Q_\alpha$ allowing for
the recoil correction (where $A_p$ is the mass number of the parent nucleus).
Fig.~\ref{figure3} shows schematically the $\alpha$ decays from the g.s. and
isomers in $^{270}$Ds. The calculated $\alpha$ decays of the two-quasineutron
$\nu 9^-$ and $\nu 10^-$ isomers to the corresponding isomers of $^{266}$Hs
have very similar $\alpha$-particle energies, giving an average value of 
11.06 MeV. This is in good agreement with the observed energy of 10.95 MeV
\cite{Hofmann01}. The $\nu 9^-$ and $\nu 10^-$ isomeric $\alpha$ decays to
the g.s. of $^{266}$Hs also have similar $\alpha$ energies with an average
value of 11.93 MeV, in agreement with the measured energy of 12.15 MeV
\cite{Hofmann01}. One also expects that the isomer can feed by
$\alpha$-emission into other states with low angular momenta.

Also from $^{270}$Ds, an observed $\alpha$-emission with $E_\alpha=11.15$ MeV
was found in coincidence with a 218 keV $\gamma$-ray signal,
and was suggested to decay to the $8^+$ member of the rotational 
g.s.b. of $^{266}$Hs \cite{Hofmann01} (the $\gamma$-ray energy 
was found to be close to the calculated energy of the transition
from the $8^+$ to $6^+$ states of the g.s.b. in $^{266}$Hs).
We calculate that the $8^+$ to $6^+$ transition
has an energy of 234 keV. The calculated $\alpha$ energies of the $\nu 9^-$
and $\nu 10^-$ decays to the $8^+$ state are 11.42 and 11.29 MeV,
respectively. In $^{270}$Ds, however, we found another low-lying $10^-$ 
state that has a two-quasiproton configuration  
$\pi\frac{11}{2}^+[615]\otimes\frac{9}{2}^-[505]$ with a calculated
excitation energy of 1.07 MeV, giving
$E_\alpha(\pi 10^-\rightarrow 8^+)=11.10$ MeV, in good agreement with the
observed value of 11.15 MeV. In $^{266}$Hs, the $\pi 10^-$ state has a
relatively high excitation energy of 2.54 MeV.

To understand the above observations, we note that the probability for
$\alpha$ decay is determined by two factors: {\sl i}) an exponential
dependence on the height of the barrier that the $\alpha$ particle needs to
penetrate and {\sl ii}) the preformation factor, characterising the ease
with which an $\alpha$ particle is formed at the surface. The latter is
proportional to the pair density at the Fermi surface\cite{Pog69},
implying a considerable reduction for two-qp states, as is the case for
high-$K$ isomers. An $\alpha$ decay between two different isomeric
configurations is in principle forbidden. Furthermore, if the
$\alpha$ particle carries a certain angular momentum, the barrier is
increased due to the centrifugal potential. The increase of the barrier
height and width reduces the $\alpha$-emission probability and hence
increases the lifetime. However, due to the exceptionally high energy of
the $\alpha$ particle for the case of superheavy nuclei, the centrifugal
barrier is of less importance, and the decay from an isomer to the g.s.
can in fact compete with the decay to the corresponding isomer.
This is consistent with the observation of the relatively long lifetimes
for the $\alpha$ decays of the isomers in $^{270}$Ds and $^{266}$Hs
\cite{Hofmann01}. Therefore, the high-$K$ isomerism can enhance the
stability of such nuclei against $\alpha$-emission. 
 
High-$K$ isomers have also been found in $^{250,256}$Fm and
$^{254}$No. In $^{250}$Fm, an 1.8-s isomer \cite{Ghi73} 
was observed at an excitation energy of 1.0 MeV \cite{Fire96}.
The $^{254}$No nucleus has a 0.28-s isomer \cite{Ghi73}
with an ${\rm energy}>0.5$ MeV \cite{Fire96}, and there is recent evidence
for associated collective structure \cite{butler}.
However, experiments have not determined spins and parities
for these two isomers. A likely configuration in $^{250}$Fm is
$\pi\frac{7}{2}^+[633]\otimes\frac{7}{2}^-[514]$
resulting in a $7^-$ state. As shown in Fig.~\ref{figure2},
the calculated excitation energy of the $7^-$ configuration is 1.01 MeV,
which agrees well with the experimental value of 1.0 MeV \cite{Fire96}.
A two-quasineutron $8^-$ configuration,  
$\nu\frac{7}{2}^+[624]\otimes\frac{9}{2}^-[734]$,
is also possible with a calculated energy of 0.97 MeV in $^{250}$Fm. 
Our calculations show that $8^-$ states exist systematically in the
$N=150$ isotones with excitation energies around 1.0 MeV.
In $^{254}$No, alternative configurations for the  observed isomer
are $\nu\frac{7}{2}^+[613]\otimes\frac{9}{2}^-[734]$ and
$\pi\frac{7}{2}^-[514]\otimes\frac{9}{2}^+[624]$
with $E_{\rm ex}^{\rm cal}=1.12$ and 1.48 MeV, respectively. The low-lying
$\nu\frac{7}{2}^+[613]\otimes\frac{9}{2}^-[734]$ states exist systematically 
in the $N=152$ isotones with calculated energies around 1.1 MeV.

Another interesting example is the $^{256}$Fm nucleus in which
a 70-ns, $7^-$ isomer was observed to undergo spontaneous
fission \cite{Hall89}. The $7^-$ isomer was populated from the $\beta$
decay of an $8^+$ isomer of odd-odd $^{256}$Es. The $^{256}$Es $8^+$ isomer
has a half-life of 7.6 hours, which is much longer than the 25 minutes of
the corresponding $1^+$ g.s. We calculate that the excitation energy of the
$7^-$ ($\pi\frac{7}{2}^+[633]\otimes \frac{7}{2}^-[514]$)
isomer in $^{256}$Fm is 1.1 MeV (see Fig.~\ref{figure2})
about 0.3 MeV less than the experimental value of 1.42 MeV \cite{Hall89}. 

\begin{figure}
\includegraphics[scale=0.45]{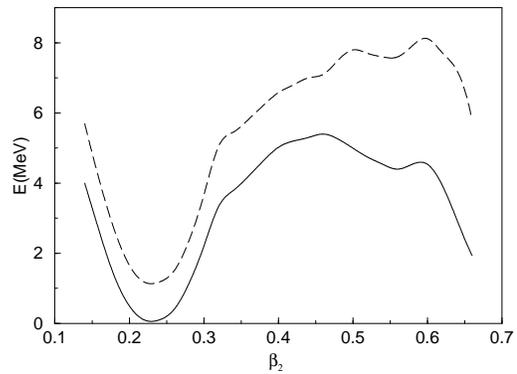}
\caption{\label{figure4}Calculated potential energies versus $\beta_2$
deformation in $^{256}$Fm. The solid and dashed lines represent the
ground state and the $\pi\frac{7}{2}^+[633]\otimes\frac{7}{2}^-[514]$ isomer,
respectively. At each $\beta_2$ point, the energy has been
minimized with respect to the $\gamma$ and $\beta_4$ deformations.}
\end{figure}

\begin{figure}
\includegraphics[scale=0.5]{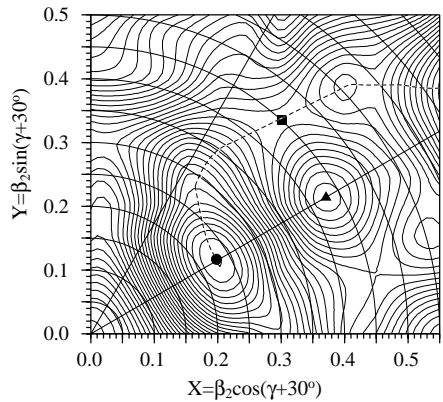}
\caption{\label{figure5}Calculated PES for the ground state of $^{256}$Fm.
The dashed line indicates the fission trajectory. The filled circle,
triangle and square are at the minimum, maximum and saddle points,
respectively. At each ($\beta_2, \gamma$) point, the energy has been
minimized with respect to the $\beta_4$ deformation.}
\end{figure}

In $^{256}$Fm, the observed partial fission half-life of the $7^-$ isomer is
0.8 ms, which is remarkably longer than the expected value of 2.5 $\mu$s
assuming no hindrance from unpaired nucleons \cite{Hall89}. 
In Ref.~\cite{Bjo80}, it has been shown that the spontaneous-fission half-life
increases considerably due to the effect of the unpaired nucleon in odd-mass
nuclei. This can be understood in terms of reduced superfluidity.
The configuration-constrained PES calculations allow us to determine 
the fission barrier for a given configuration. Fig.~\ref{figure4} shows
the configuration-constrained fission barrier for the $7^-$ isomer in
$^{256}$Fm. (It needs to be mentioned that an isomer might undergo a more
complex dynamical process during fission, without keeping a fixed
configuration.) The increase of the fission barrier
in both height and width implies an increase in the fission lifetime.
In the calculations of fission barriers, the inclusion of
the non-axial $\gamma$ deformation is important because the $\gamma$ degree
of freedom can significantly affect the shape of the fission barrier.
In $^{256}$Fm, for example, the non-axial PES calculations result in 
the barrier heights being lowered by 2.4 and 2.6 MeV for the g.s. and
$7^-$ isomer, respectively, compared with axial calculations.
Fig.~\ref{figure5} displays the calculated PES of the g.s. in $^{256}$Fm,
with the fission trajectory marked. The $7^-$ isomer has a similar fission
trajectory.
 
Although the production of high-spin states in superheavy nuclei is
generally inhibited by their increased fission probability,
this does not apply when the angular momentum is non-collective and
aligned with the symmetry axis, as for high-K states. Therefore,
there should be significant production of high-K isomers in fusion reactions.
Two-qp excitations can also appear in odd-mass and odd-odd superheavy 
nuclei, resulting in three- and four-qp states, respectively. These have
not yet been considered in detail. Furthermore, the present investigation
is limited to deformed superheavy nuclei. However, the angular-momentum
effects should also apply to spherical superheavy nuclei.

In summary, we have investigated multi-qp states in even-even
superheavy nuclei using configuration-constrained PES calculations.
Low-energy high-$K$ deformed configurations are predicted to be
present in the vicinity of the Fermi surface, resulting in the
formation of high-spin isomeric states. The high-spin isomerism
increases the barrier to fission and decreases the probability
of $\alpha$-particle emission. For a given class of superheavy nuclei
with sufficiently short lifetimes, we have discussed an inversion
of stability, in that the excited states can be longer lived than the
corresponding ground states. These effects result in increased survival
probabilities of superheavy nuclei. Therefore, it can be stated
that via the population of high-$K$ states in experiments,
one may be able to extend the nuclear chart further into the island
of superheavy nuclei. The energy calculations of multi-qp states and
their $\alpha$ decays indicate that the single-particle orbits of
the deformed Woods-Saxon potential apply also to the superheavy mass
region. Further studies of high-$K$ states will provide essential 
tests, not only of our specific isomer predictions, but also of mean-field 
models in general.  

This work was supported by the Chinese Major State Basic Research Development
Program G2000077400, the Chinese National Natural Science Foundations
(Grant Nos. 10175002 and 10047001), the Doctorial Foundation of Chinese 
Ministry of Education (20030001088), the Knowledge Innovation Project of
Chinese Academy of Sciences (KJCX2-SW-N02), the Swedish Science Research
Council(VR), the UK Royal Society, and the UK Science and Engineering 
Research Council.


\begin{thebibliography}{99}
\bibitem{Hofmann00} S. Hofmann and G. M{\" u}nzenberg, Rev. Mod. Phys.
{\bf 72}, 733 (2000). 
\bibitem{Oganessian99} Yu.Ts. Oganessian, {\it et al.}, Phys. Rev. C
{\bf 63}, 011301(R) (2001); Phys. Rev. C {\bf 69}, 021601(R) (2004).
\bibitem{Nilsson68} S.G. Nilsson {\it et al.}, Nucl. Phys. {\bf A115},
545 (1968).
\bibitem{Cwiok96} S. {\' C}wiok, J.~Dobaczewski, P.-H.~Heenen, P.~Magierski,
W.~ Nazarewicz, Nucl. Phys. {\bf A611}, 211 (1996).
\bibitem{Smol97} R. Smola{\' n}czuk, Phys. Rev. C {\bf 56}, 812 (1997).
\bibitem{Rutz97} K. Rutz {\it et al.}, Phys. Rev. C {\bf 56}, 238 (1997).
\bibitem{Bender99} M. Bender, K.~Rutz, P.G.~Reinhard, J.A.~Maruhn,
W.~Greiner, Phys. Rev. C {\bf 60}, 034304 (1999).
\bibitem{Herzberg01} R.-D. Herzberg {\it et al.}, Phys. Rev. C {\bf 65},
014303 (2001).
\bibitem{Reiter99} P. Reiter {\it et al.}, Phys. Rev. Lett.
{\bf 82}, 509 (1999).
\bibitem{Fire96} {\it Table of Isotopes}, 8th ed., edited by
R.B.~Firestone and V.S.~Shirley (Wiley, New York, 1996) Vol. 2.
\bibitem{Sob01} A. Sobiczewski, I. Muntian, Z. Patyk, Phys. Rev. C
{\bf 63}, 034306 (2001).
\bibitem{Cwiok99} S. {\' C}wiok, W. Nazarewicz and P.H. Heenen
 Phys. Rev. Lett. {\bf 83}, 1108 (1999).
\bibitem{Xu98} F.R. Xu, P.M. Walker, J.A. Sheikh and R. Wyss,
Phys. Lett. {\bf B435}, 257 (1998).
\bibitem{Naz85} W. Nazarewicz, J.~Dudek, R.~Bengtsson, T.~Bengtsson,
I.~Ragnarsson, Nucl. Phys. {\bf A435}, 397 (1985).
\bibitem{Sat94}W. Satu{\l}a, R. Wyss and P. Magierski,
Nucl Phys. {\bf A578} 45 (1994). 
\bibitem{Mol92} P. M{\"o}ller and J.R. Nix, Nucl. Phys. {\bf A536},
20 (1992).
\bibitem{Mye66} W.D. Myers and W.J. Swiatecki, Nucl. Phys. {\bf 81},
1 (1966).
\bibitem{Hofmann01} S. Hofmann {\it et al.}, Eur. Phys. J. {\bf A10},
5 (2001).
\bibitem{Pog69} J.K. Poggenburk, H.J. Mang, and J.O. Rasmussen,
Phys. Rev. {\bf181}, 1697 (1969).
\bibitem{Ghi73} A. Ghiorso, K. Eskola, P.~Eskola, and M.~Nurmia,
Phys. Rev. C {\bf 7}, 2032 (1973).
\bibitem{butler} P.A. Butler {\it et al.}, Phys. Rev. Lett.
{\bf 89}, 202501 (2002).
\bibitem{Hall89} H.L. Hall {\it et al.}, Phys. Rev. C {\bf 39},
1866 (1989).
\bibitem{Bjo80} S. Bjornholm and J.E. Lynn, Rev. Mod. Phys. {\bf 52},
725 (1980).
\end{thebibliography}
\end{document}